\documentstyle[twocolumn]{article}

\def\bar{\overline}

\def\d{\delta}
\def\a{\alpha}
\def\b{\beta}
\def\n{\nu}
\def\m{\mu}
\def\k{\kappa}
\def\e{\epsilon}

\def\th{\theta}

\def\bar{\overline}
\def\l{\lambda}

\def\eV{{\rm eV}}
\begin{document}
\twocolumn[
\centerline{\Large\bf Solar Neutrino Solutions in Non-Abelian Flavor
Symmetry}
\vskip 0.5 cm
\centerline{\large\bf Morimitsu Tanimoto}
\vskip 0.3 cm
\centerline{\it Department of Physics, Niigata University, 
 Niigata, JAPAN}
\vskip 0.5 cm
\centerline{( Talk at ICHEP2000, July 27-August 2, 2000, Osaka, Japan )}
\vskip 0.8 cm

We have studied the large mixing angle
MSW solution for the solar neutrinos in  the non-abelian flavor
symmetry. We predict the MNS mixing matrix taking account of the symmetry
breakings.  
\vskip 1  cm  ]
\noindent
{\large\bf 1  LMA-MSW Solution  in  Non-Abelian Flavor  Symmetry}
\vskip 0.3 cm
 Recent  data in S-Kam favor
  the large mixing  angle MSW (LMA-MSW) solution.
   How does  one  get the   LMA-MSW solution as well as the maximal
mixing of  the atmospheric neutrinos in theory?
  It is not easy to reproduce the nearly bi-maximal mixings
with LMA-MSW solution in GUT models\cite{NY,NS,AB}.

The non-abelian flavor symmetry
 $S_{3L}\times S_{3R}$ or  
$O_{3L}\times O_{3R}$ 
leads to the LMA-MSW solution naturally\cite{TWY,T}.
The  mass matrices are
\begin{eqnarray}
 M_E\propto
             \left (\matrix{1 & 1 & 1 \cr
			    1 & 1 & 1 \cr
                            1 & 1 & 1 \cr  } \right) , \
 M_\n\propto\left( \matrix{1 & 0 & 0 \cr
                                0 & 1 & 0 \cr
                                0 & 0 & 1 \cr  } \right) .
\nonumber
\end{eqnarray}
\noindent
The orthogonal matrix diagonalizes  $M_E$ is
\begin{eqnarray}
F=\left( \matrix{1/\sqrt 2 & 1/\sqrt 6 & 1/\sqrt 3 \cr
                   -1/\sqrt 2 & 1/\sqrt 6 & 1/\sqrt 3 \cr
                           0 & -2/\sqrt 6 & 1/\sqrt 3 \cr } \right). 
\end{eqnarray}
\noindent
The MNS mixing matrix is given as $U_\nu\simeq F^T$, and so 
we  predict
$\sin^2 2\th_\odot=1$ and $\sin^2 2\th_{\rm atm}=8/9$ in the
symmetric limit.

In this talk, we discuss  masses and flavor mixings of  quarks/leptons
 in the non-abelian flavor symmetry with the SU(5) GUT\cite{TWY}.
    We consider $O(3)_{\bf 5^*}\times O(3)_{\bf 10} \times Z_6$  symmetry.
Our scenario for  fermion  masses is

\noindent
$\bullet$ Neutrinos have degenerate masses.

\noindent
$\bullet$ Quarks/charged-leptons are massless.

\noindent
$\bullet$ Symmetry breakings  give  $\Delta m^2$ and other fermion masses.

\vskip 0.3 cm
\noindent
{\large\bf 2 $\bf O(3)_{\bf 5^*}\times O(3)_{\bf 10} \times Z_6$ Symmetry}
\vskip 0.3 cm
Quarks and leptons belong to   $\bf 5^*$ and $\bf 10$ of the $SU(5)$ GUT and
   $\bf 3$ of the $O(3)$ symmetry.
 Higgs $H$ ($\bar H$) belong to  $\bf 5$ ($\bf  5^*$) of the $SU(5)$ 
 and  $\bf 1$ of the $O(3)$.
Then, 
neutrinos have the $O(3)_{\bf 5^*} \times O(3)_{\bf 10}$ invariant mass term
\begin{eqnarray}
  \frac{<H>^2}{\Lambda} \nu_L \nu_L \ .
\end{eqnarray}
 The $Z_6$ symmetry  forbids  
    $\psi_{\bf 10}({\bf 3}) \psi_{\bf 10}({\bf 3}) H$,
which gives  degenerate up-quark masses\cite{TWY}.

  The flavor symmetry is broken explicitly  by
$\Sigma^{(i)}_{\bf 5^*}({\bf 5},  {\bf 1}) , \  
  \Sigma^{(i)}_{\bf 10}({\bf 1},  {\bf 5})  \  (i=1, 2)$, 
which  transform as the symmetric traceless tensor {\bf 5}'s of  $O(3)$.
Dimentionless breaking parameters are given as 
\begin{eqnarray}
 \sigma^{(1)}_{\bf 10,\  \bf 5^*}\equiv \frac{\Sigma^{(1)}_{\bf 10,\ 5^*}}{M_f} =
  \left( \matrix{1 & 0 & 0 \cr
        0 & 1 & 0 \cr  0 & 0 & -2  \cr  } \right )  \delta_{\bf 10,\ 5^*} ,
\nonumber
\end{eqnarray}	
 \begin{eqnarray}
 \sigma^{(2)}_{\bf 10,\  \bf 5^*}\equiv 
\frac{\Sigma^{(2)}_{\bf 10,\  \bf 5^*}}{M_f} =
  \left( \matrix{1 & 0 & 0 \cr
        0 & -1 & 0 \cr  0 & 0 & 0 \cr  } \right )  \e_{\bf 10,\ 5^*}  .
\nonumber
 \end{eqnarray}	
\noindent
   Neutrinos get Majorana masses from a superpotential
\begin{eqnarray}
  W =\frac{H^2}{\Lambda}\ell ( {\bf 1}+\a_i\sigma^{(i)}_{5^*}) \ell \  ,
\end{eqnarray}
\noindent
which yields a diagonal neutrino mass matrix.
In order to get the charged lepton masses,
  we introduce  $O(3)_{\bf 5^*}$-triplet $\phi_{\bf 5^*}({\bf 3},{\bf
1})$ and $O(3)_{\bf 10}$-triplet  $\phi_{\bf 10}({\bf 1},{\bf 3})$. 
 These VEV's  are determined by the superpotential
 \begin{eqnarray}
W = Z_{\bf 5^*}(\phi^2_{\bf 5^*} - 3v_{\bf 5^*}^2) + 
        Z_{\bf 10} (\phi^2_{\bf 10} - 3v_{\bf 10}^2) \nonumber \\
 +X_{\bf 5^*} (a_{(i)} \phi_{\bf 5^*}\sigma_{\bf 5^*}^{(i)}\phi_{\bf 5^*} )  +  X_{\bf 10} ( a'_{(i)}\phi_{\bf 10} \sigma_{\bf 10}^{(i)} \phi_{\bf 10} ) 
\nonumber \\
+ Y_{\bf 5^*} ( b_{(i)}\phi_{\bf 5^*} \sigma_{\bf 5^*}^{(i)}\phi_{\bf 5^*})  
+  Y_{\bf 10} ( b'_{(i)} \phi_{\bf 10} \sigma_{\bf 10}^{(i)} \phi_{\bf 10})
\nonumber
\end{eqnarray}	
\noindent where
 $Z_{\bf 10, \ \bf 5^*}$, $X_{\bf 10, \ \bf 5^*}$,
 $Y_{\bf 10, \ \bf 5^*}$ are all singlets 
 of $O(3)_{\bf 5^*} \times O(3)_{\bf 10} $.
\noindent
Minimizing the potential, we get 
 \begin{eqnarray}
 <\phi_{\bf 5^*}>\equiv \left (\matrix{1\cr 1 \cr 1 \cr} \right )v_{\bf 5^*},
 \ 
 <\phi_{\bf 10}>\equiv \left ( \matrix{1\cr 1 \cr 1 \cr} \right )v_{\bf 10}.
\nonumber
 \end{eqnarray}
  \noindent
   Masses of charged leptons arise from a superpotential
 \begin{eqnarray}
      W = \frac{\k_E}{M_f^2} (\bar e\phi_{\bf 10}) (\phi_{\bf 5^*} \ell)
  \bar H   ,
   \end{eqnarray}
  \noindent which is the 
   realization of   "Democratic  Mass Matrix",
\begin{eqnarray}
   M_E \propto \left ( \frac{v_{\bf 5^*} v_{\bf 10}}{M_f^2}\right )
    \left (\matrix{1 & 1 & 1 \cr } \right )
     \left (\matrix{ 1  \cr  1 \cr  1 \cr }  \right ). 
  \end{eqnarray}
\noindent Adding the superpotential containing 
 the flavor symmetry  breaking parameters  
  $\sigma_{\bf 5^*,\ \bf 10}^{(i)}$,
  we get the charged lepton mass matrix:
\begin{eqnarray}
&& M^H_E\equiv  F^T M_E  F = 
  \k_E \left ( \frac{v_{\bf 5^*} v_{\bf 10}}{M_f^2}\right )  <\bar H> 
\nonumber \\
&& \times  \left ( \matrix{\e_{\bf 5^*} \e_{\bf 10} & \e_{\bf 10}\d_{\bf 5^*} 
                          &  \e_{\bf 10} \cr
  \e_{\bf 5^*} \d_{\bf 10} &  \d_{\bf 5^*}
 \d_{\bf 10}  &  \d_{\bf 10}  \cr 
 \e_{\bf 5^*}  & \d_{\bf 5^*}  & 3  \cr  
	 } \right )  , 
\end{eqnarray}
 \noindent  
in  which  order one coefficients are omitted.  The mass ratios  are
given as 
\begin{eqnarray}
   \frac{m_\m}{m_\tau} \simeq {\cal O}(\d_{\bf 5^*}\d_{\bf 10}), \quad 
  \frac{m_e}{m_\tau} \simeq {\cal O}(\e_{\bf 5^*}\e_{\bf 10}).
\nonumber
 \end{eqnarray}
\noindent
The quark/lepton masses and mixings fix  
\begin{eqnarray}
  \d_{\bf 10} \simeq \l^2  , \ \e_{\bf 10} \simeq \l^3\sim \l^4,\
 \d_{\bf 5^*}\simeq \l  , \ 
   \e_{\bf 5^*} \simeq \l^2 .  \nonumber
 \end{eqnarray}
\vskip 0.3 cm
\noindent
{\large\bf 3 Neutrino Masses and  Mixings}
\vskip 0.3 cm
 Neutrino masses are given as
\begin{eqnarray}
m_1\simeq &&c_\mu(1+\a_1\d_{\bf 5^*}+\a_2\e_{\bf 5^*}) ,
\nonumber\\ 
m_2 \simeq &&c_\mu (1+\a_1\d_{\bf 5^*}-\a_2\e_{\bf 5^*}),
 \nonumber\\
m_3 \simeq && c_\mu(1-2 \a_1\d_{\bf 5^*}), 
\quad c_\mu=\frac{<H>^2}{\Lambda}  \nonumber
\end{eqnarray}
\noindent which leads to (${\rm with} \ \d_{\bf 5^*}\simeq \l  , \ \
   \e_{\bf 5^*} \simeq \l^2 $)
\begin{eqnarray}
 \left | \frac{\Delta m_{21}^2}{\Delta m_{32}^2}\right |
=&& \frac{2}{3}\frac{\a_2\e_{\bf 5^*}}{\a_1\d_{\bf 5^*}}
 \frac{1+\a_2\e_{\bf 5^*}}{1-\frac{1}{2}\a_1\d_{\bf 5^*}}
\simeq  \l^2 \sim \l .\nonumber
\end{eqnarray}
\noindent
Putting  $\Delta m^2_{32}= 3\times  10^{-3} \eV^2$, 
we predict
 $\Delta m_{21}^2 \simeq ({\rm factor})\times 10^{-4} \eV^2 $,
which is  consistent with the LMA-MSW solution.
Flavor  mixings come from the  charge lepton mass matrix since
the neutrino one is  diagonal.
The charged lepton mass matrix is diagonalized by $V_R^\dagger  M_E^H
V_L$, in which
\begin{eqnarray}
 V_L^\dagger \simeq \left( \matrix{1 & \l &  \l^2 \cr
               - \l & 1 & \l \cr   -\l^2  & -\l  & 1\cr } \right) .
\end{eqnarray}
\noindent The neutrino mixing matrix is given 
 by $ V_L^\dagger  F^T$. We predict 
\begin{eqnarray}
&&\sin^2 2\th_\odot=(1-\frac{4}{3}\l^2)^2\simeq 0.87 
  \nonumber \\
&&\sin^2 2\th_{\rm atm}=
   \frac{8}{9}(1-\l^2)(1+\frac{1}{\sqrt{2}}\l-2 \l^2)^2
\nonumber \\ &&\hskip 2 cm \simeq 0.95 \nonumber \\
&&\left | U_{e3} \right | = 
\frac{2}{\sqrt{6}}\l (1-\frac{1}{\sqrt{2}}\l)\simeq 0.14.        
\end{eqnarray}

\vskip 0.3 cm
\noindent
{\large\bf 4 Summary}
\vskip 0.3 cm
It is remarked that:

\noindent
$\bullet$  The solar neutrino mixing $\sin^2 2\th_\odot$ deviates from
the maximal mixing ($\sim 0.87$).

\noindent
$\bullet$ The atmospheric neutrino mixing $\sin^2 2\th_{\rm atm}$ deviates from
$8/9$  depending phase of $\l$.

\noindent
$\bullet$  $\bf U_{e3}$ is near  to the experimental bound of CHOOZ 
  ($\bf \leq 0.16$) .

Neutrino masses are degenerated within a factor 2.
For example, we get
\noindent
$m_1\simeq 0.030  \eV, \  m_2\simeq 0.033 \eV, \ 
    m_3 \simeq 0.058 \eV$,
which is consistent with $\b\b_{0\nu}$ decay bound.


\end{document}